\documentclass[preprint2]{aastex}

\shorttitle{High-energy pulsations from PSR J1741$-$2054}
\shortauthors{Marelli et al.}

\begin{document}

\title{X and $\gamma$-ray pulsations of the nearby radio-faint PSR J1741$-$2054}

\author{M. Marelli\textsuperscript{1}, A. Belfiore\textsuperscript{1,4},
P. Saz Parkinson\textsuperscript{4,7}, P. Caraveo\textsuperscript{1,2}, A. De Luca\textsuperscript{1,2}, C. Sarazin\textsuperscript{5},
D. Salvetti\textsuperscript{1,2,3}, G.R. Sivakoff\textsuperscript{5,6}, F. Camilo\textsuperscript{8,9}}

\footnote{INAF - Istituto di Astrofisica Spaziale e Fisica Cosmica Milano, via E. Bassini 15, 20133 Milano, Italy\\
\textsuperscript{2}Istituto Nazionale di Fisica Nucleare, Sezione di Pavia, Via Bassi 6, I-27100 Pavia (Italy)\\
\textsuperscript{3}Universit\`a degli Studi di Pavia, Strada Nuova 65, 27100 Pavia, Italy\\
\textsuperscript{4}Santa Cruz Institute for Particle Physics, University of California, Santa Cruz, CA 95064\\
\textsuperscript{5}Department of Astronomy, University of Virginia, P.O. Box 400325, Charlottesville, VA 22904-4325, USA\\
\textsuperscript{6}Department of Physics, University of Alberta, CCIS 4-183 Edmonton, AB T6G 2E1, Canada\\
\textsuperscript{7}Department of Physics, The University of Hong Kong, Pokfulam Road, Hong Kong\\
\textsuperscript{8}Columbia Astrophysics Laboratory, Columbia University, New York, NY 10027, USA\\
\textsuperscript{9}Arecibo Observatory, HC3 Box 53995, Arecibo, PR 00612, USA}
\email{marelli@lambrate.inaf.it}
\author{M. Marelli\altaffilmark{1}, A. Belfiore\altaffilmark{1,4},
P. Saz Parkinson\altaffilmark{4,9}, P. Caraveo\altaffilmark{1,2}, A. De Luca\altaffilmark{1,2}, C. Sarazin
\altaffilmark{5}, D. Salvetti\altaffilmark{1,2,3}, G.R. Sivakoff\altaffilmark{5,6}, F. Camilo\altaffilmark{7,8}}

\altaffiltext{1}{INAF - Istituto di Astrofisica Spaziale e Fisica Cosmica Milano, via E. Bassini 15, 20133 Milano, Italy}
\altaffiltext{2}{Istituto Nazionale di Fisica Nucleare, Sezione di Pavia, 
Via Bassi 6, I-27100 Pavia (Italy)}
\altaffiltext{3}{Universit\`a degli Studi di Pavia, Strada Nuova 65, 27100 Pavia, Italy}
\altaffiltext{4}{Santa Cruz Institute for Particle Physics, University of California, Santa Cruz, CA 95064}
\altaffiltext{5}{Department of Astronomy, University of Virginia, P.O. Box 400325, Charlottesville, VA 22904-4325, USA}
\altaffiltext{6}{Department of Physics, University of Alberta, CCIS 4-183 Edmonton, AB T6G 2E1, Canada}
\altaffiltext{7}{Columbia Astrophysics Laboratory, Columbia University, New York, NY 10027, USA}
\altaffiltext{8}{Arecibo Observatory, HC3 Box 53995, Arecibo, PR 00612, USA}
\altaffiltext{9}{Department of Physics, The University of Hong Kong, Pokfulam Road, Hong Kong}
\email{marelli@lambrate.inaf.it}

\begin{abstract}

We report the results of a deep {\it XMM-Newton} observation of the radio-faint $\gamma$-ray pulsar J1741$-$2054
and its nebula together with the analysis of five years of {\it Fermi} Large Area Telescope (LAT) data.
The X-ray spectrum of the pulsar is consistent with an absorbed power law plus a blackbody,
originating at least partly from the neutron star cooling.
The nebular emission is consistent with that of a synchrotron pulsar wind nebula, with hints of
spatial spectral variation.
We extended the available {\it Fermi} LAT ephemeris and folded the $\gamma$-ray and X-ray data.
We detected X-ray pulsations from the neutron star: both the thermal and non-thermal components are $\sim$35-40\%
pulsed, with phase-aligned maxima. A sinusoid fits the thermal folded profile well.
A 10-bin phase-resolved analysis of the X-ray emission shows softening of the non-thermal spectrum during the on-pulse phases.
The radio, X-ray and $\gamma$-ray light curves are single-peaked, not phase-aligned, with the X-ray peak trailing the $\gamma$-ray
peak by over half a rotation. 
Spectral considerations suggest that the most probable pulsar distance is in the 0.3-1.0 kpc range, in agreement with the radio dispersion
measure.

\end{abstract}


\section{Introduction} \label{intro}

The launch of the {\it Fermi $\gamma$-ray Space Telescope} offered the first opportunity to study a
sizeable population of $\gamma$-ray pulsars.
The {\it Fermi} Large Area Telescope \citep[LAT,][]{atw09} has discovered pulsed $\gamma$-ray signals
from more than 150 objects \citep{abd13}, revolutionizing our view of them and
giving birth to new high-energy pulsar sub-families, such as millisecond \citep[see e.g.][]{abd09a,ran11,kei11,esp13} and radio-quiet
$\gamma$-ray pulsars \citep[see e.g.][]{abd09,saz10,ple13}, as numerous as the classic family of young, radio-loud pulsars \citep{car13}. 
The wealth of detections confirms the importance of the $\gamma$-ray channel
in the overall energy budget of rotation-powered pulsars and paves the way to
understanding the three-dimensional structure and electrodynamics of neutron star magnetospheres.
Indeed, radio and $\gamma$-ray light curves contain a great deal of useful
information about pulsar emission processes
\citep[see e.g.][]{wat11,pie12,pie14}, confirming that models with emission originating at high
altitudes in the magnetosphere \citep[e.g. outer and slot-gap,][]{che86,har04} are favored over models with near-surface emission
\citep[e.g. polar cap,][]{har13}.

Fitting $\gamma$-ray and radio light curves simultaneously is a promising way to constrain pulsar geometric parameters  \citep[e.g.,][]{pie14}. 
Using the information in the (magnetospheric) non-thermal pulsar X-ray light curves
could further improve the approach, adding another piece to the pulsar
emission puzzle. This approach could localize the emitting region(s)
responsible for the non-thermal pulsed X-ray emission with respect to the
high altitude gamma-ray emitting region.\\
Few X-ray light curves have been exploited for modeling magnetospheric emission, compared to $\gamma$-ray profiles.
This is largely due to the lack of high-quality X-ray light curves, primarily due to the occasional
and non-targeted observational efforts.
At this time, $\sim$60 out of 77 young pulsars in the second {\it Fermi} LAT pulsar catalog \citep[2PC,][]{abd13} 
have been detected in X-rays \citep{bec09,mar11,mar12,abd13}, but X-ray pulsations have been detected from fewer than half of them.
Only nine {\it Fermi} pulsars have both the high X-ray fluxes and the
long dedicated X-ray observations needed to disentangle the thermal and the non-thermal pulsations.
Only five of these (Crab, Vela, Geminga, PSR J0659+1414 and PSR J1057$-$5226) have been 
characterized by a multi-bin phase-resolved X-ray spectral analysis \citep{del05,man07,wei11}.
Of these only the Crab \citep[and possibly Geminga,][]{jac05} yielded a non-thermal light curve with a
photon index varying with phase, a behavior that is still puzzling \citep{har08,tan08,hir08}.

With the notable exception of the Crab among young pulsars, the multiwavelength behavior of isolated neutron stars is 
complex, with radio, optical, X-ray and $\gamma$-ray light curves 
usually unaligned, pointing to different emitting regions in the pulsar magnetosphere.
The rich phenomenology, in particular including the X-ray information, has not yet been
fully exploited for modeling the radiation processes of pulsars, leaving a number of questions unsolved. 

Here we report the results of a deep {\it XMM-Newton} observation intended for phase-resolved X-ray spectral
analysis of the {\it Fermi} pulsar J1741$-$2054 (hereafter, J1741).
The middle-aged J1741 ($\tau_c$ = 390 kyr) was discovered in a blind pulsation search of a {\it Fermi} LAT point source \citep{abd09}. 
For a moment of inertia $I = 10^{45}$ g cm$^2$,
its period $P$ = 413 ms and period derivative $\dot{P}$ = $1.7\times10^{-14}$ s s$^{-1}$
give a spin-down energy loss rate $\dot{E}$ = 9$\times10^{33}$ erg s$^{-1}$, 
clearly on the low side of the $\gamma$-ray pulsar distribution \citep[see Figure 1 in ][]{abd13}.
The pulsar was then detected in archival Parkes radio observations, showing 
a remarkably low dispersion measure, DM = 4.7 pc cm$^{-3}$ \citep{cam09}.
The Galactic electron density model of \citet{cor02} yields a distance of $\sim$0.38 kpc,
making J1741 one of the closest $\gamma$-ray pulsars known. 
At this distance, the low observed 1400 MHz radio flux density, S$_{1.4}$ $\sim$ 160 $\mu$Jy, 
makes it the faintest radio pulsar known.

At the position obtained from LAT timing analysis \citep{ray11},
\citet{abd09}, \citet{rom10} and \citet{mar11} found the X-ray counterpart using both {\it Swift} and {\it Chandra} data. 
The {\it Chandra} observation also revealed diffuse, faint X-ray emission due to a pulsar wind nebula (PWN) 
trail extending some 2$'$ at position angle P.A. = 45$^{\circ}$ $\pm$ 5$^{\circ}$ (north through east).
This extended structure was also associated with a 20$''$ long H$\alpha$ bow shock.
Accurate bow shock modelling by \citet{rom10} suggests a pulsar velocity
v$_p$ $\sim$ 150 km s$^{-1}$ directed 15$^{\circ}$ $\pm$ 10$^{\circ}$ out of the plane of the sky.

\section{X-ray Observations} \label{obs}

Our deep {\it XMM-Newton} observation of J1741 started on 2013 February 28 at 19:50:39 UT
and lasted 70.9 ks. The PN camera \citep{str01} of the European Photon Imaging Camera (EPIC) was operating
in Small Window mode (time resolution of 5.6 ms over a 4$'$ $\times$ 4$'$ field of view, FOV),
while the Metal Oxide Semi-conductor (MOS) detectors \citep{tur01} were set in Full Frame mode (2.6 s time resolution and
a 15$'$ radius FOV). The thin optical filters were used for the PN and MOS cameras.
For our analysis, we used the {\it XMM-Newton} Science Analysis Software (SAS) v13.0. 
To fully characterize both the pulsar and the nebula, we also used the available {\it Chandra}
Advanced CCD Imaging Spectrometer (ACIS)
\citep{gar03} observation of the field, performed on 2010 May 21 and
lasting 48.8 ks \citep[these data were included in][]{rom10}.
To better characterize the pulsar, we also took advantage of the $\sim$300 ks of data collected as part of the Cycle 14 {\it Chandra}
Visionary Project $``$A Legacy Study of the Relativistic Shocks of PWNe$"$ by R. Romani.
We retrieved $``$level 1$"$ data from the {\it Chandra} Science Archive and used the {\it Chandra}
Interactive Analysis of Observation (CIAO) software v.4.3.
For all the datasets, we followed the standard
data processing approach, using the same procedures reported in \citet{mar13}.

\section{$\gamma$-ray analysis} \label{gamma}

The {\it Fermi} LAT dataset we used to extend $\gamma$-ray ephemeris of J1741 spans five years from 2008
August 4 to 2013 August 4.
P7REP {\tt Source} class events were selected with reconstructed energies from 0.1 to 100 GeV
from an area within 20$^{\circ}$ of the source position.
We excluded $\gamma$ rays collected when the LAT was not in nominal science operations mode, 
when the spacecraft rocking angle exceeded 52$^{\circ}$, when the Sun was within 5$^{\circ}$
of the pulsar position, and those with measured zenith angles $>100^{\circ}$, 
to reduce contamination by residual $\gamma$ rays from the bright limb of the Earth.
We performed a binned maximum likelihood analysis, as reported in \citet{abd13}.
We used the {\it Fermi} Science tools v09r32p04, instrument response functions P7REP\_SOURCE\_V15,
the Galactic and isotropic models obtained by the LAT collaboration from the analysis of four years of data.
The analysis tools, instrument response functions, and diffuse emission models are available from the {\it Fermi} Science Support Center,
\url{http://fermi.gsfc.nasa.gov/ssc}.
The source models were taken from the two-year catalog \citep{nol12} and 2PC.
Post-fit spatial residuals did not reveal the need for any additional sources,
beyond those in the two-year catalog, in our model of the region.
The pulsar $\gamma$-ray spectrum is consistent with a power law with an exponential cutoff
with a photon index $\Gamma$ = 1.10$\pm$0.10 and cutoff energy E$_c$ = 0.92$\pm$0.06 GeV.
These results are in full agreement with those in 2PC.

The {\it Fermi} Science Tool {\tt gtsrcprob} combines the spectral results with
the LAT's energy-dependent point-spread function (PSF) to assign each event its probability of 
coming from the pulsar \citep{ker11}. 
We used only barycentered events with a probability higher than $0.01$ for the following timing analysis.
The rotation ephemeris used in 2PC spans only three years:
we extended it, using a weighted Markov chain Monte Carlo algorithm \citep[MCMC, see e.g.][]{wan13}.
Adding six months of data in each iteration, we re-evaluated the timing solution using the weighted H-test \citep[see e.g.][]{{dej10}}.
The highest H-value of 4782.59 (20 harmonics) for five years of data was found for $F_0$=2.41720384698 Hz, $F_1$=-9.93133e-14,
$F_2$=-5.924e-25 and $F_3$=5.75e-32, with epoch zero at MJD 55631.0002. \footnote{The rotation ephemeris is available at
http://fermi.gsfc.nasa.gov/ssc/data/access/lat/ephems/}\\
With this ephemeris we assigned a rotational phase to each $\gamma$-ray event and filled a 100-bin $>$0.1 GeV phase histogram, 
with bin uncertainties using the photon weights, following \citet{abd13} (see Section~\ref{timing}).

We performed a $\gamma$-ray phase-resolved spectral analysis on the five-year dataset. 
For 20 phase bins we re-ran the binned spectral analysis, leaving only one or more of the pulsar parameters (normalization,
photon index and energy cutoff) free to vary.
The Test Statistic value \citep[see e.g. ][]{mat96} does not vary significanlty by freeing one or two of the parameters
so that the addition of the additional degrees of freedom are not justified.
Moreover, with both the photon index and the cutoff energy free, their fit with a constant is acceptable (null hypothesis probability 
$nhp$=6$\times10^{-4}$ and 6$\times10^{-3}$ for the photon index and cutoff respectively).
Thus, we found no indication of spectral variation as a function of the pulsar phase.

\section{X-ray Imaging and Source Detection} \label{ima}

Following the method reported in \citet{mar13}, we detected and selected the AGN-like serendipitous
sources in the {\it XMM-Newton} and {\it Chandra} FOVs (see Figure~\ref{fig1}). 
The spectra of these sources were fitted together, 
linking their hydrogen column densities N$_H$ to assess the average Galactic absorption column.
The resulting value, N$_H^{gal}$ = (4.54$\pm$0.36) $\times$ 10$^{21}$ cm$^{-2}$ (90\% confidence error), 
is slightly greater than the value of 3$\times$10$^{21}$ cm$^{-2}$ obtained from the 21 cm HI sky survey of \citet{kal05}.
This points to the absence of thick molecular clouds in the {\it XMM-Newton} FOV.
Since N$_H^{gal}$ obtained from X-ray absorption probes all types of nuclei along the line of sight,
whereas the latter value samples only atomic hydrogen, this observed value of N$_H$ is to be expected \citep{dic90}.

Apart from the pulsar itself, the main feature apparent in the PN FOV is the PWN emission.
To find the best extraction regions for the pulsar and the nebula, as well as to better evaluate the PWN shape,
we created exposure-corrected radial (around the pulsar) and linear (parallel and orthogonal to
the main axis of the nebula) brightness profiles in the 0.3-10 keV energy range.
The radial profile is consistent ($nhp$ = 0.14) with the nominal instrument PSF \citep{rea04}
only up to 40$''$ from the pulsar due to the relative faintness of the nebula.
The linear profile, with 1$'$ width centered on the pulsar
and orthogonal to the main axis of the nebula, is also consistent ($nhp$ = 0.009) with the
PSF, proving the absence of a detectable nebular component in the orthogonal region.
The presence of the nebula in the profile parallel to the main axis is apparent,
extending up to $\sim2'$ from the pulsar and with flux decreasing with distance \citep[apparent in Figure~\ref{fig1}, see also][]{rom10}.
From the deep {\it Chandra} observations the unusual, shattered shape of the tail-like nebula is apparent,
appearing to be divided into three blobs.

\section{Phase-integrated spectral analysis} \label{spint}

Using a 25$''$ radius extraction circle around the pulsar, 
we obtained 9794, 2983 and 3073 counts in the 0.3-10 keV energy range in
the PN and the two MOS detectors, respectively, with a background contribution of less than 4\%. 
We also used 19380 pulsar counts from the {\it Chandra} observations. 
For the spectral analysis, we followed \citet{mar13}.
To better constrain the column density, we fitted the nebular spectra together with the pulsar spectrum.
We obtained 1508, 557 and 533 counts in the 0.3-10 keV energy range in
the PN and the two MOS detectors respectively, with a background contribution of 40\% and 25\%.
We obtained 5633 nebular counts, with a 30\% background contribution from the {\it Chandra} observations.

One-component models are not statistically acceptable for the pulsar spectrum 
($\chi^2_{r}$=2.2, $nhp$ $\sim0$ for an absorbed power law and $\chi^2_{r}$=11.6, $nhp$=0 for an absorbed blackbody, 
where the $\chi^2$ has been reduced for 781 degrees of freedom, or $``$dof$"$).
The pulsar spectrum is consistent ($\chi^2_{r}$=1.18, $nhp$=$2\times10^{-3}$, 779 dof) 
with a combination of a power law component with a photon
index of $\Gamma$ = 2.68$\pm$0.04 (90\% confidence errors) and a blackbody
with a temperature of T = (7.07$\pm$0.19) $\times$ $10^5$ K, and an
emitting radius R = (5.39$_{-0.71}^{+0.81})\times d_{380}$ km
(where $d_{380}$ is the distance of the pulsar in units of $380$ pc, derived from the dispersion measure
of the pulsar using the NE2001 electron model of the Galaxy, \citet{cam09}), absorbed by a column
density of N$_H$ = (1.21$\pm$0.01) $\times$ 10$^{21}$ cm$^{-2}$, about one fourth of the Galactic value.
The PWN spectrum is consistent with an absorbed power law with $\Gamma$ = 1.74$\pm$0.07.
The unabsorbed fluxes of the non-thermal and thermal components of the pulsar spectrum are
5.47$\pm$0.13 $\times10^{-13}$ and 7.63$\pm$0.19 $\times10^{-13}$ erg cm$^{-2}$ s$^{-1}$, respectively. The unabsorbed nebular flux is
1.40$\pm$0.09 $\times10^{-13}$ erg cm$^{-2}$ s$^{-1}$.\\
We note that a three component model is not statistically compelling (for a power law plus double
blackbody we obtain $\chi^2_{r}$=1.18, $nhp$=$3\times10^{-3}$, 777 dof).
We also note that a composite non-thermal plus a magnetized neutron star atmosphere model
({\tt nsa} in XSpec; assuming a NS with a radius of 12 km, mass of 1.4 M$_{\odot}$, and a surface
magnetic field of 10$^{13}$ G) gives a very poor fit.
We obtain a $\chi^2_{r}$=1.22, $nhp$=$7\times10^{-5}$, 779 dof, with a lower emitting
temperature of (3.02$\pm$0.12) $\times$ $10^5$ K.

To study the possible spectral variation of the nebula with angular distance from the pulsar,
we divided the nebula into three different regions on the basis of their angular separation from the pulsar.
We chose to equally divide the main axis of the ellipsoidal nebular region, in order to
consider the three blobs. The different PSFs, as well as different exposure maps, of the two X-ray instruments
have been taken into account.
All the parameters of the spectra from the corresponding regions in {\it Chandra} and {\it XMM-Newton} observations
were linked, also freezing the absorption column to the previously fitted value.
We see marginal evidence for a spectral variation (in fact, linking the 
photon index for the three spectra we obtain $\chi^2_{r}$=1.45, $nhp$=6.4$\times10^{-5}$,184 dof),
with $\Gamma_{near}$=1.72$\pm$0.09 (90\% confidence errors),
$\Gamma_{med}$=1.90$\pm$0.10 and $\Gamma_{far}$=1.69$\pm$0.09.

\section{Pulsar timing analysis} \label{timing}

For the timing analysis, PATTERN selection was performed as by 
\citet{mar13} and the X-ray photon arrival times were barycentered 
to the {\it Chandra} source position (R.A,decl.(J2000) 17:41:57.28, -20:54:11.8) from \citet{rom10},
which is consistent with the $\gamma$-ray timing position.
We then phase-folded the X-ray photons using our {\it Fermi} LAT timing solution,
contemporaneous with our {\it XMM-Newton} data set. 
We extracted 11507 PN events in the 0.15-10 keV energy range from the 25$''$ circle centered on the pulsar.
We repeated the exercise for photons in different energy ranges.

Pulsations with $>20\sigma$ significance appear in the 0.15-10 keV energy range 
(H-test=585; tail probability = 0, and $\chi^2_{r}$=29, 19 dof, according to a $\chi^2$ test on the folded curve
testing a constant model). 
The profile has a single peak, lagging the radio peak by 0.6 in phase.
Light curves for different energy ranges are shown in Figure ~\ref{fig4}, 
phase-aligned with the $\gamma$-ray and radio profiles.   
Using the phase-averaged composite spectrum, we evaluated the background contribution,
the pulsar blackbody, and the power law components in different energy ranges,
as in \citet{car10}.
We obtained a background-subtracted pulsed fraction of 36.1$\pm$1.5\% in the 0.3-10 keV energy range.
Dividing the energy range into low energy (0.3-0.7 keV, where the blackbody accounts for $\sim$60\% of the counts)
and high-energy bands (0.7-10 keV, where the power law accounts for $\sim$90\% of the counts)
we obtained pulsed fractions of 35.8$\pm$1.5\% and 36.4$\pm$1.5\%, respectively.
We know the pulsed fraction in each band and the percentage of
blackbody, power-law and background contributions, so that we can get the
net pulsed fraction of the two spectral components.
The pulsar's power law is pulsed at $\sim$38\% and the blackbody
component at $\sim$36\%.
While at lower energies the curve is quasi-sinusoidal 
(a fit with a sinusoid gives $\chi^2_{r}$=1.31 , 17 dof, $nhp$=0.17)
due to the blackbody contribution, 
the profile at higher energies is not compatible with a sinusoid ($\chi^2_{r}$=2.70 , 17 dof, $nhp$=10$^{-4}$).
The correlation of the power-law photon index with its normalization (see e.g. Figure~\ref{fig6}) 
somewhat biases the normalization profile in Figure~\ref{fig5}, while
the counts profiles in Figure~\ref{fig4} are better suited for comparing their 
shapes.
The peak positions at low (0.3-0.7 keV) and high (0.7-10 keV) X-ray energies are separated by less than $0.1$ in phase, 
but they are offset by about a half-rotation from the main $\gamma$-ray pulse and the radio pulse
(following \citet{abd13}, the radio pulse leads the $\gamma$-ray pulse by 0.074$\pm$0.006 in phase).

\section{Phase-resolved pulsar analysis} \label{spres}

To search for possible variation of the X-ray spectral parameters with rotational phase, 
we first fitted the on- and off-pulse spectra. 
We define the on-pulse interval to be between phase 0.3 and 0.8 (values obtained from a fit with a double-step model),
and the remaining phase bins as off-pulse.
Fitting only the power-law counts (E$>$0.7 keV), we fixed the column density and the photon
index to the phase-averaged values, obtaining an acceptable $nhp$=0.047 ($\chi^2_{r}$=1.21, 143 dof). 
Freeing both the normalization and photon indices we obtain an improved fit ($nhp$=0.22, $\chi^2_{r}$=1.09, 141 dof).
An f-test \citep{bev69} shows that the probability for a chance improvement is 2.6$\times10^{-4}$,
pointing to a significant effect when both the power law normalization and photon index are free to vary.\\
Next we fixed the power law values obtained in the previous fit and we evaluated the blackbody component in the 0.3-0.7 keV energy band.
Thawing only the blackbody normalization, we obtain a $nhp$=0.74 ($\chi^2_{r}$=0.93, 128 dof).
A variation of the blackbody temperature with pulsar phase is not statistically compelling;
however, the high pulsed fraction at low energies implies significant pulsation of the blackbody normalization.

The blackbody normalization and power-law normalization and photon index vary with pulsar phase. We divided our
dataset into 10 phase bins, each with width 0.1 and containing $\sim$1000 counts,
and fitted the obtained spectra fixing the column density and linking
blackbody temperatures. We obtained the curves reported in Figure~\ref{fig5}, with the
photon index gradually increasing with phase and the normalizations following the lightcurve, with
no statistically significant differences between blackbody and powerlaw peaks. 
The variation in photon index is also shown by the confidence contours for the 10 phase bins,
plotted in Figure~\ref{fig6} for the plane of the index and the power law normalization.
The variation with blackbody normalization is similar.

\section{Discussion and Conclusions}

By analyzing the new {\it XMM-Newton} observation of PSR J1741$-$2054, we fully characterized the
high-energy emission of this nearby middle-aged radio-faint pulsar.\\
Its nebular emission is typical of pulsar wind nebulae, both for its non-thermal spectrum,
and for the flux decrease with distance along the tail emanating from the pulsar.
A hint of nebular spectral variation with distance from the pulsar is detected.
The shattered shape of the nebula is peculiar and a deeper analysis of the new {\it Chandra} data is needed to better understand
this unusual behavior. These observations could also provide a measurement of the pulsar proper motion.
Such an analysis is beyond the scope of this work.

Modeling the X-ray spectrum of the pulsar requires a composite model, summing thermal and non-thermal components.
Both components are $\sim$35-40\% pulsed, with single-peaked light curves and maxima phase-aligned to within
$0.1$ in phase.
While the thermal light curve is compatible with a sinusoid, the non-thermal profile has a sharper peak.
The best-fitting thermal spectrum yields a temperature ($\sim7\times10^5$ K), compatible with the
theoretical expectations for the cooling of a 390 kyr-old pulsar \citep{pon09}.
We note that the best fit temperature of the {\tt nsa} model ($\sim3\times10^5$ K) is 
below the theoretical expectations, further disfavoring such a model.
A pulsed component from thermal cooling has already been noted for several pulsars \citep{car04,del05,man07},
and that of PSR J1057$-$5226 has a similar pulsed component. Such pulsations
can be ascribed to a dependence of the observed emitting area on the line of sight.
\citet{hal93} describe a magnetospheric $``$blanket$"$ caused
by cyclotron resonance scattering off the plasma in the magnetosphere that could
screen the thermally emitting surface during specific phase intervals, depending on the magnetic field configuration
and viewing geometry. 
Anisotropic heat transfer from the pulsar interior can also explain flux variation across the neutron star surface \citep{gre83}.
If the thermal component is due to the cooling of the entire neutron star surface (12 km radius), from the thermal normalization
in the best fit spectrum (that depends only on the pulsar distance and the emitting radius)
we can derive the pulsar distance to be $\sim$850 pc, with 3$\sigma$ limits of $\sim$650 and $\sim$1100 pc \citep[e.g.,][]{hal07}.
On the other hand, thermal emission from polar caps heated due to downstreaming of $e^{\pm}$ \citep{har02},
as seen in the case of PSR J0007+7303 in the CTA 1 supernova remnant, 
is expected to be generated from much smaller regions
($<$100 m, based on a simple $``$centered$"$ dipole magnetic field geometry, \citet{del05,mar13}),
which would imply an unrealistic pulsar distance smaller than 10 pc.
Distances lower than a few hundred parsecs are greatly disfavored due to
the non-negligible value of the column density (about one fourth of the total Galactic column density in this direction).
We note that a distance of 850 pc would result in a 110\% $\gamma$-ray efficiency, defined as $L_\gamma/\dot E$,
with $L_\gamma$ the luminosity above 100 MeV. The distance range cited above implies an efficiency range of 60 to 180\%.
Similarly, the X-ray efficiency would be $\sim$1.2\% (0.7 to 2.1\%).
Such an unrealistic $\gamma$-ray efficiency could be explained by a beaming factor f$_{\Omega}$ less than 1 (as defined in 2PC),
or by a moment of inertia larger than 10$^{45}$ g cm$^2$.

A 10-bin phase-resolved X-ray spectral analysis reveals variations in the X-ray photon index, 
in addition to the phase-varying normalizations of the two spectral components, 
with a softer spectrum during the on-pulse phases (Figure \ref{fig5}).
It is difficult to compare the behavior of J1741 with that of other {\it Fermi} pulsars, 
since the Crab is the only one for which such variation was detected,
and the Crab's non-thermal spectrum becomes softer in the primary-pulse maximum and
harder during the bridge between the two maxima \citep{wei11}.
Although some models have been developed to explain the Crab's optical-to-$\gamma$-ray behavior
\citep[see e.g.][]{har08,tan08,hir08}, the physics behind the pulsar's X-ray photon index variations is still unclear. 

Moreover, unlike the Crab, the $\gamma$-ray, X-ray and radio peaks of PSR J1741$-$2054
are not aligned, pointing to a clear difference in the geometry and/or
altitude above the neutron star surface of the different emitting regions. 
Indeed, the $\gamma$-ray and X-ray peaks are phase-offset by roughly a half rotation, 
as is also seen in PSRs J0007+7303, J1057$-$5226 and J0659+1414 \citep{del05,car10}.
Although these differences are expected from different
models for the radio and $\gamma$-ray bands \citep{abd13}, no model is able to account for the offset between $\gamma$-ray
and X-ray peaks.
The alignment between the thermal and non-thermal X-rays (seen also in other pulsars, e.g. J0659+1414 and J1057$-$5226),
as well as the phase lag with the $\gamma$-ray emission coming
from the outer magnetosphere, can suggest that the non-thermal emission is generated in
a region near the pulsar poles (e.g. in a polar cap emission model). Also, the low X-ray luminosity of radio-quiet pulsars
in the X-ray band \citep{mar11} suggests that the radio and X-ray emission regions may be in close proximity.
The origin of the phase lag between the radio
and X-ray light curves is unclear. A comprehensive study of pulsar high-energy
light curves and phase-resolved spectra will be crucial to understanding the X-ray emission mechanisms and geometry.

\begin{figure*}
\centering
\includegraphics[angle=0,width=15cm]{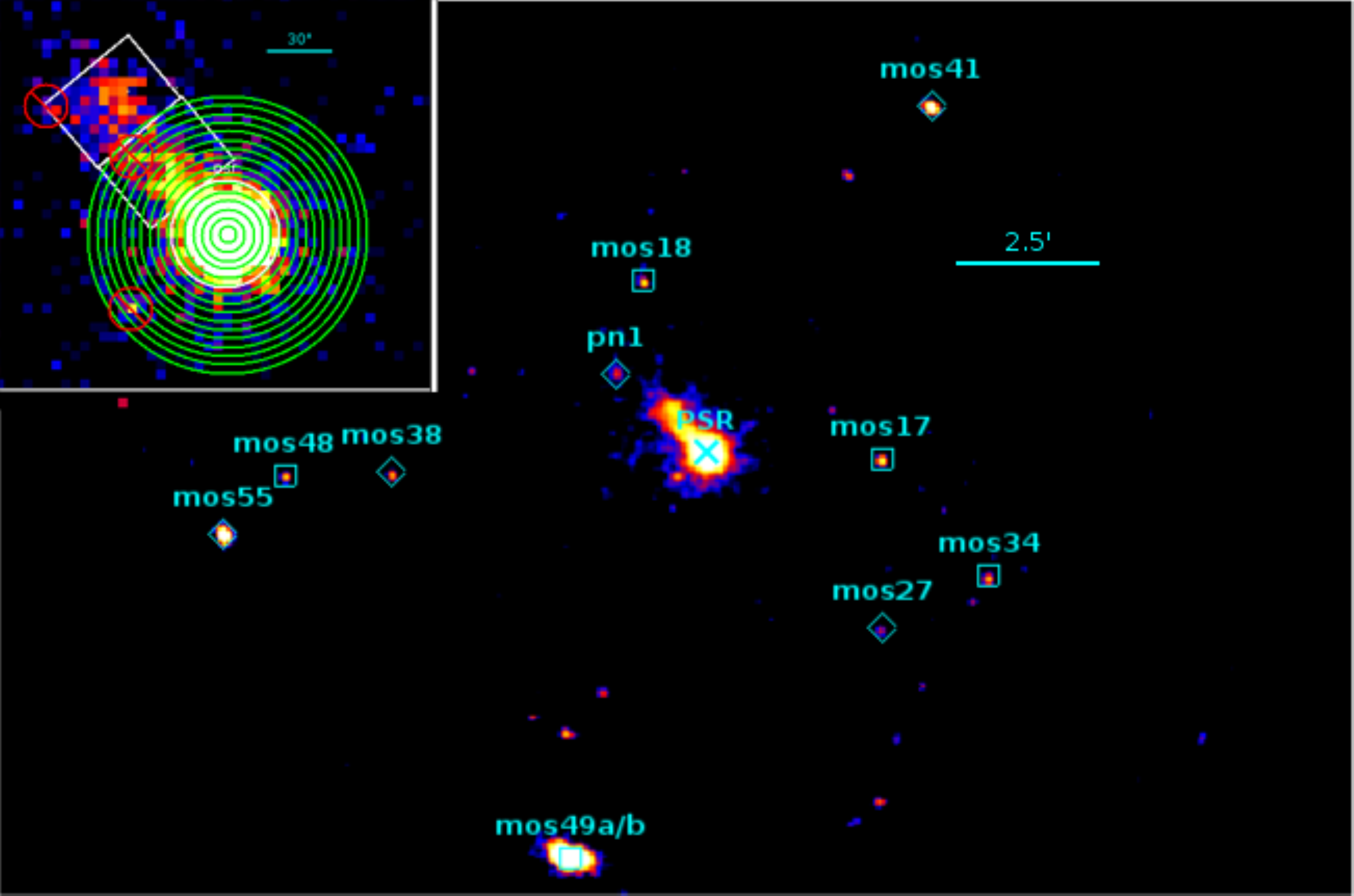}
\protect\caption{{\footnotesize Combined exposure-corrected 0.3-6 keV FOV images of the three {\it XMM-Newton} cameras. We applied
a Gaussian filter with a kernel radius of 3$''$. Cyan symbols mark the sources detected with $>6\sigma$ confidence
and with more than 225 detected counts that we analyzed
to constrain the value of the Galactic absorption column.
The squares mark the AGN-like sources, diamonds other field sources and the cross marks the pulsar.
{\it Upper-left panel}: Expanded image of the pulsar and its tail. Green annuli mark the regions we used for the PSF analysis
(with the exception of the nebular region); from white regions we extracted photons to build the pulsar and nebular spectra;
red regions mark the pointlike sources we excluded from analysis.}}
\end{figure*}

\begin{figure*}
\centering
\includegraphics[angle=0,width=15cm]{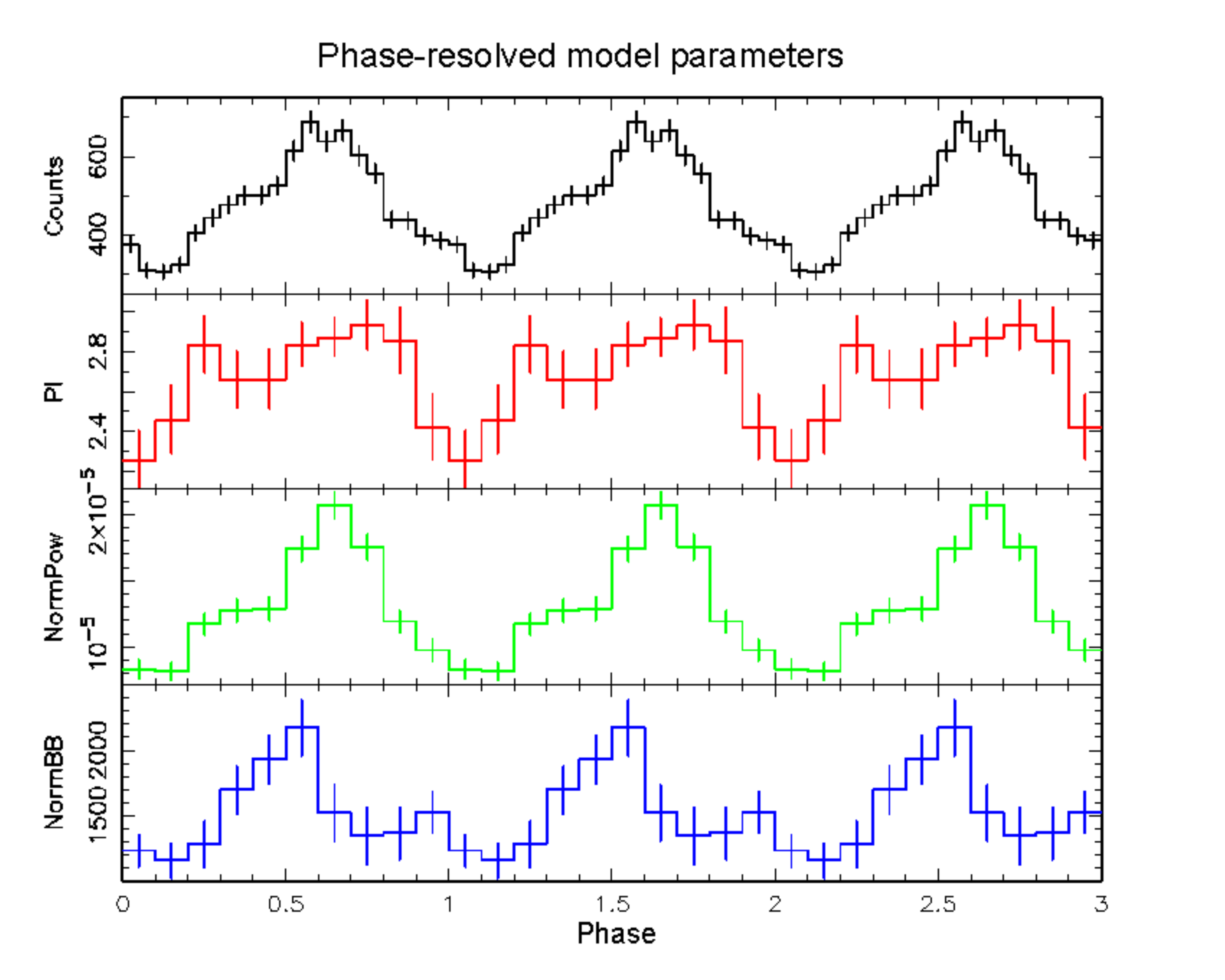}
\protect\caption{{\footnotesize Best-fit phase-resolved X-ray spectral parameters of PSR J1741$-$2054,
plotted as a function of the pulsar phase, defined as in Figure 3. As discussed in Section~\ref{spres},
the power law (panel 3 from the top) and blackbody (panel 4) normalizations and
the photon index (panel 2) evolve throughout the pulsar phase.}}
\end{figure*}

\begin{figure*}
\centering
\includegraphics[angle=0,width=15cm]{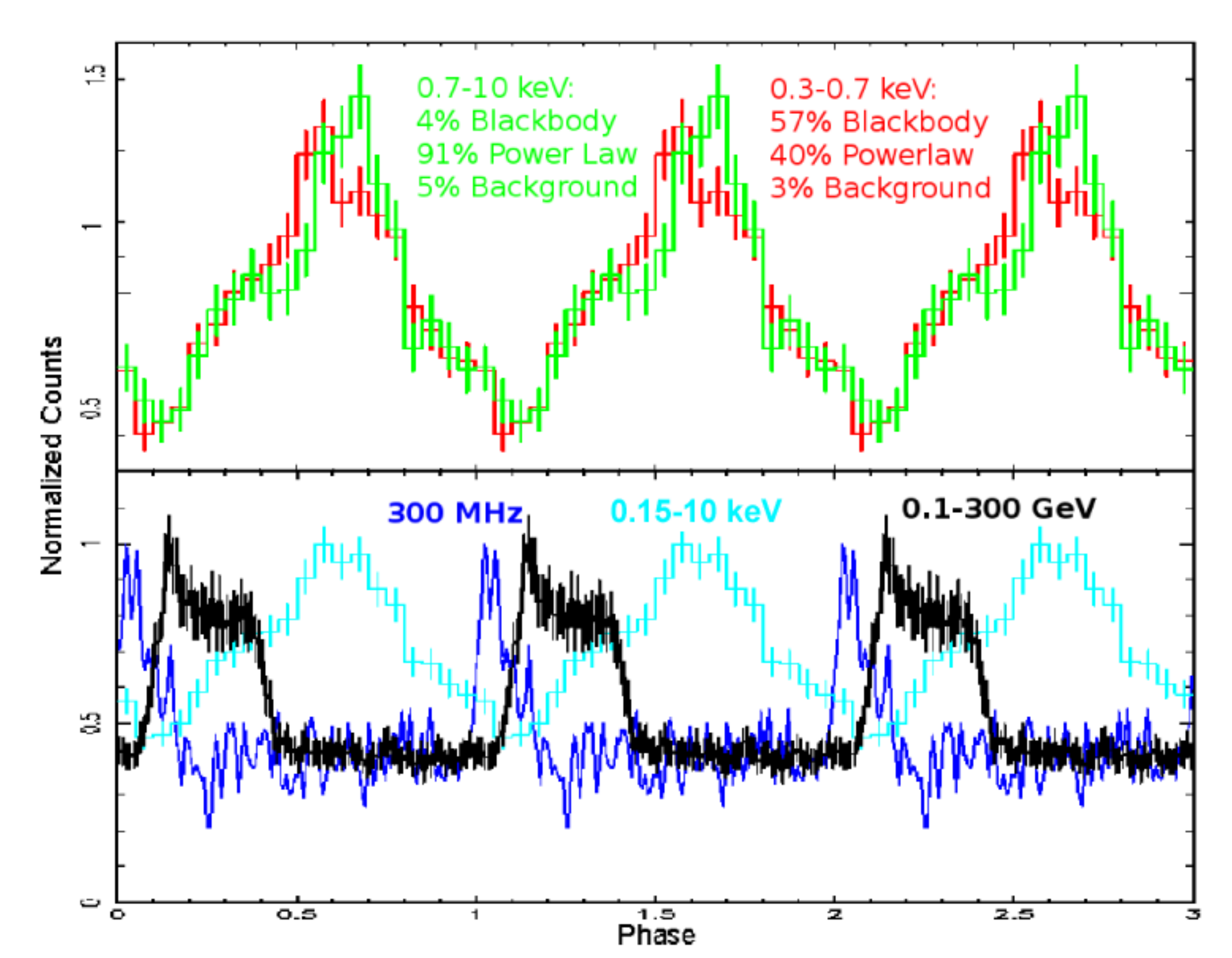}
\protect\caption{{\footnotesize {\it Top Panel}: EPIC/pn folded light curves in different energy ranges.
X-ray photon phases were computed according 
to the {\it Fermi} LAT ephemeris overlapping
with the {\it XMM-Newton} dataset, with selection as in Section 6.
The red curve contains photons in the 0.3-0.7 keV energy range and
the green one in the 0.7-10 keV range.
The curves have been renormalized by dividing each bin by N$_{counts}$/N$_{bins}$, where N$_{counts}$ is the 
total number of events in the energy range and N$_{bins}$ the number of bins (1$\sigma$ errors are shown).
{\it Bottom Panel}: Phased radio, X-ray, and $\gamma$-ray light curves of PSR J1741$-$2054.
The 300 MHz radio light curve, shown in blue, comes from the Green Bank Observatory \citep{cam09}.
The 100-bin $\gamma$-ray curve, shown in black, contains all the 5-year {\it Fermi} LAT weighted counts
with energies $>$100 MeV. 
The 20-bin X-ray curve is shown in cyan. All the curves have been renormalized to
have the highest bin value equal to 1 (1$\sigma$ errors are shown).}}
\end{figure*}

\begin{figure*}
\centering
\includegraphics[angle=0,width=15cm]{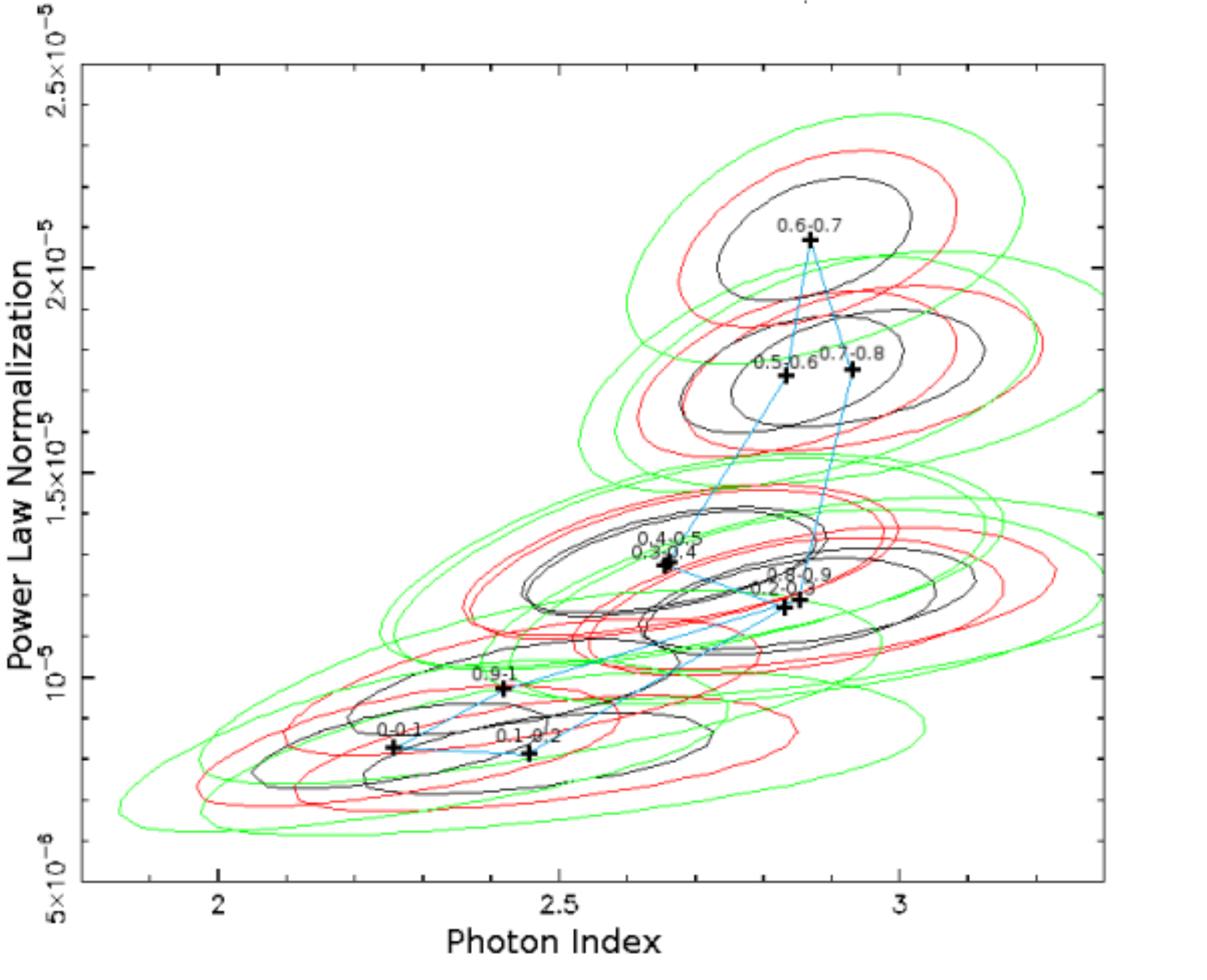}
\protect\caption{{\footnotesize Confidence contours for the 10-bin X-ray phase-resolved
analysis of PSR J1741$-$2054, showing the pulsar photon index
and the power law normalization. Black contours are at the 1$\sigma$ confidence level, red at 90\% and green at 3$\sigma$.
Blue lines follow the pulsar phase.}}
\end{figure*}

\acknowledgments

This work was supported by the ASI-INAF contract I/037/12/0,
art.22 L.240/2010 for the project $''$Calibrazione ed Analisi del satallite NuSTAR$"$.
Craig Sarazin was supported in part by NASA ADAP grant NNX13AE64G.
Gregory Sivakoff is supported by an NSERC discovery grant.\\
The \textit{Fermi} LAT Collaboration acknowledges generous ongoing support
from a number of agencies and institutes that have supported both the
development and the operation of the LAT as well as scientific data analysis.
These include the National Aeronautics and Space Administration and the
Department of Energy in the United States, the Commissariat \`a l'Energie Atomique
and the Centre National de la Recherche Scientifique / Institut National de Physique
Nucl\'eaire et de Physique des Particules in France, the Agenzia Spaziale Italiana
and the Istituto Nazionale di Fisica Nucleare in Italy, the Ministry of Education,
Culture, Sports, Science and Technology (MEXT), High Energy Accelerator Research
Organization (KEK) and Japan Aerospace Exploration Agency (JAXA) in Japan, and
the K.~A.~Wallenberg Foundation, the Swedish Research Council and the
Swedish National Space Board in Sweden.
Additional support for science analysis during the operations phase is gratefully
acknowledged from the Istituto Nazionale di Astrofisica in Italy and the Centre National d'\'Etudes Spatiales in France.\\
A special thanks to David A. Smith and Tyrel Johnson for reviewing the paper.

Facilities: CXO (ACIS), XMM (EPIC), Fermi (LAT).

\clearpage

\end{document}